\documentclass[%
 reprint,
nofootinbib,
 amsmath,amssymb,
 aps,
 prd,
]{revtex4-2}

\usepackage{graphicx}
\usepackage{bm}
\usepackage{xcolor}
\usepackage{array}
\usepackage{makecell}

\newcommand{\Mpl}{M_{\mathrm{Pl}}}

\newcommand{\eps}{\varepsilon}

\newcommand{\Rthree}{{}^{(3)}R}
\newcommand{\VD}{V_{\mathcal D}}
\newcommand{\avglat}[1]{\left\langle #1\right\rangle_{\mathrm{lat}}}
\newcommand{\avgV}[1]{\left\langle #1\right\rangle_V}
\newcommand{\grad}{\bm{\nabla}}

\newcommand{\mpl}{M_{\mathrm{Pl}}}

\newcolumntype{C}[1]{>{\centering\arraybackslash}p{#1}}

\usepackage{hyperref}
\hypersetup{
  colorlinks=true,
  linkcolor=blue,
  citecolor=blue,
  urlcolor=blue
}

\begin{document}

\title{A note on oscillons: non-negligible metric fluctuations during preheating}

\author{Daniel del~Corral}
\email{daniel.corral.martinez@uj.edu.pl}
\affiliation{Faculty of Physics, Astronomy and Applied Computer Science, Jagiellonian University, 30-348 Krakow, Poland}

\author{Paolo Gondolo}
\email{paolo.gondolo@utah.edu}
\affiliation{Department of Physics and Astronomy, University of Utah,
Salt Lake City, UT 84112, USA}
\affiliation{School of Science, Institute of Science Tokyo,
2-12-1 Ookayama, Meguro-ku, Tokyo 152-8551, Japan}

\author{K.~Sravan~Kumar}
\email{sravan.kumar@port.ac.uk}
\affiliation{Institute of Cosmology and Gravitation, University of Portsmouth,
Dennis Sciama Building, Burnaby Road, Portsmouth, PO1~3FX, United Kingdom}

\author{Jo\~ao Marto}
\email{jmarto@ubi.pt}
\affiliation{Departamento de F\'isica, Centro de Matem\'atica e Aplicações (CMA-UBI), Universidade da Beira Interior,
Rua Marqu\^es D'\'Avila e Bolama 6200-001, Covilh\~a, Portugal}

\date{\today}


\begin{abstract}
The dominant lattice approach to post-inflationary preheating and oscillon formation evolves an inhomogeneous inflaton field on a spatially homogeneous Friedmann-Lema\^itre-Robertson-Walker background, whose expansion is sourced by a volume-averaged energy density, while metric perturbations are neglected. We examine the consistency of this approximation with General Relativity. Using the linearized Einstein constraints, we show that the suppression of the Bardeen potential, characteristic of slow-roll inflation, disappears during preheating: once the first slow-roll parameter becomes of order unity, the metric and inflaton fluctuations enter at the same perturbative order. We further show that the scalar-field equation evolved in the fixed-FLRW prescription omits leading-order contributions generated by metric fluctuations. We then derive the proper-volume average of the ADM Hamiltonian constraint and show that it does not reduce to the Friedmann equation used in lattice simulations: the exact averaged constraint contains additional contributions from the spatial curvature, the variance of the local expansion, the shear, and metric corrections to the local energy density. Finally, we illustrate numerically, for Starobinsky and $\alpha$-attractor models, that the metric contribution becomes comparable to the scalar-field contribution during the amplification stage. These results directly affect the standard lattice description of oscillon formation and motivate numerical relativity as a consistent framework for the nonlinear preheating problem.
\end{abstract}


\maketitle


\section{Introduction}

Several lattice codes are designed to study the evolution of interacting scalar fields in an expanding universe. \texttt{LATTICEEASY} \cite{Felder:2000hq} is one of the first, dating back to the year 2000. It is based on finite-difference methods to compute the spatial derivatives of the scalar fields that evolve on a fixed, homogeneous, and isotropic flat Friedmann-Lema\^itre-Robertson-Walker (FLRW) background, therefore ignoring metric fluctuations. Many improvements to this code have come over the years, such as \texttt{CLUSTEREASY} \cite{Felder:2007kat}, which supports parallel computations based on MPI libraries, \texttt{DEFROST} \cite{Frolov:2008hy}, with some speed and accuracy improvements, and \texttt{CUDAEASY} \cite{Sainio:2009hm} and \texttt{PyCOOL} \cite{Sainio:2012mw}, GPU-accelerated versions. Other approaches employ pseudo-spectral methods, such as \texttt{PSpectRe} \cite{Easther:2010qz} and \texttt{GABE} \cite{Child:2013ria}, or Lattice Gauge Field Theory techniques, such as \texttt{GFiRe} \cite{Lozanov:2019jff}. All these codes are based on the same fundamental implementation of \texttt{LATTICEEASY}, neglecting metric fluctuations and relying on a volume-averaged energy density that is not consistent with a proper averaging scheme. In some physical contexts, such as the preheating stage, the oscillating scalar field at the bottom of its potential induces parametric resonances \cite{del-Corral:2023apl,del-Corral:2024vcm,Martin:2019nuw,Martin:2020fgl,Jedamzik:2010dq,Ballesteros:2024eee,del-Corral:2025fca,del-Corral:2025lcp,Hertzberg:2014iza,Hertzberg:2014jza,Sfakianakis:2018lzf} that effectively amplify the scalar perturbations and therefore should not be neglected. 

Examples of applications of lattice codes include the production of gravitational waves. In 2006, the authors of \cite{Easther:2006vd,Easther:2007vj} used the results from \texttt{LATTICEEASY} to compute the gravitational-wave spectrum sourced by scalar fields, setting a precedent for lattice codes. After that, several codes started to include the computation of gravitational-wave backgrounds as one of their outputs. Also based on the original implementation of \texttt{LATTICEEASY}, some examples of these kinds of codes are \texttt{HLattice V2.0} \cite{Huang:2011gf}, which includes a sixth-order symplectic integrator and evolves the scalar fields in a perturbed FLRW background\footnote{\label{footnote}The \texttt{HLattice V2.0} code \cite{Huang:2011gf} includes metric perturbations through second order in the Einstein--Hilbert action, but, as noted by its authors, does not include the cubic interaction responsible for the scalar-induced tensor source. As a result, preheating gravitational-wave calculations based on \texttt{HLattice}, including the $\alpha$-attractor analysis of Ref.~\cite{Bhoonah:2020oov}, omit this leading contribution. Including the corresponding metric-induced source modifies the predicted gravitational-wave spectrum \cite{del-Corral:2025fzz}.}, or the recent one, $\mathcal{C}$\texttt{osmo}$\mathcal{L}$\texttt{attice} \cite{Figueroa:2021yhd,Figueroa:2020rrl,Figueroa:2023xmq,Baeza-Ballesteros:2025tme}, which goes up to tenth-order but evolves the fields in a fixed homogeneous FLRW background. These codes again ignore metric fluctuations and, consequently, do not properly compute the gravitational waves associated with the interacting scalar fields. Another example is the formation of oscillons \cite{Bogolyubsky:1976yu,Bogolyubsky:1976pw,Gleiser1994,Copeland1995}, which is the main focus of this paper due to their relevance in the context of lattice codes and the numerous works derived from the (incorrect) methodology used to study them.

The contents of this paper are organized as follows. In Sec.~\ref{sec:oscillons}, we review the formation of oscillons during preheating. Sec.~\ref{sec:caveats} explores the caveats in the lattice approach and why it is inconsistent with a proper, ADM-like, averaging scheme. In Sec.~\ref{sec:numerical-relativity}, we highlight numerical relativity as a way forward to properly include metric perturbations. Conclusions are given in Sec.~\ref{sec:conclusions}.

\section{Overview of oscillon approach and preheating}
\label{sec:oscillons}

One of the most explored effects produced by interacting scalar fields, particularly during the preheating phase, is the formation of oscillons. These objects constitute pseudo-stable, localized, and long-lived configurations of scalar fields. They were first realised in field theory \cite{Bogolyubsky:1976yu,Bogolyubsky:1976pw,Gleiser1994,Copeland1995} before appearing in cosmological scenarios, and they are still being studied in other contexts such as ultralight scalar dark matter \cite{Turner:1983he,Press:1989id} or axion dark matter \cite{Kolb:1993hw,Kolb:1994fi}. In this work, we want to clarify what we call the \textit{preheating-oscillon paradigm in the lattice approach}, a narrower and more recent framework in which oscillons are produced from the fragmentation of the inflaton scalar field condensate and claimed to dominate the energy density of the universe for several e-folds of expansion \cite{Amin:2011hj,Mahbub:2023faw,Shafi:2024jig}, which, in turn, would source a stochastic background of gravitational waves and also leave an imprint in several cosmological observables \cite{Zhou:2013tsa,Lozanov:2019ylm,Antusch:2016con,Bhoonah:2020oov}. This paradigm was established in 2010 for 1+1D \cite{Amin:2010xe} and later extended to three dimensions \cite{Amin:2010dc}. 

Several of these works employ the lattice codes mentioned above to study the formation and evolution of oscillons. However, as stated above, the codes consistently exclude metric fluctuations and, therefore, should not be trusted in these particular cases. This is the reason we frame it as a \textit{paradigm}, and we believe that it has its origin in the original implementation of \texttt{LATTICEEASY}, which has then been imprinted in the consequent codes. In Ref.~\cite{Amin:2011hj}, for example, the scalar field is evolved nonlinearly in three spatial dimensions, whereas the gravitational sector is restricted to a homogeneous FLRW background whose expansion is sourced by spatially averaged matter quantities. The local metric response to the growing scalar inhomogeneities is therefore omitted, based on the assumption that metric perturbations remain dynamically negligible during resonance.

This is the general observation we have seen so far, and that applies to the codes and works listed above, except \texttt{HLattice V2.0} (see footnote \ref{footnote}). The main objective of this work is to highlight that in some scenarios in which metric fluctuations are not small \cite{del-Corral:2023apl,del-Corral:2024vcm,Martin:2019nuw,Martin:2020fgl,Jedamzik:2010dq,Ballesteros:2024eee,del-Corral:2025fca,del-Corral:2025lcp,Hertzberg:2014iza,Hertzberg:2014jza,Sfakianakis:2018lzf}, the use of these approaches should be used with care, but even if the metric fluctuations are small, the volume-averaged energy density considered in the lattice approaches is not taken consistently, as we show below. This point warrants an important clarification, as we do not intend to diminish previous work. Rather, we aim to emphasise regimes in which metric perturbations can be as important as, or even dominate over, scalar field perturbations, so that these codes can be improved to account for this issue. We also emphasise that the structural objection of this Letter is to a specific class of codes and the methodology they employ, not to numerical methods in cosmology. Also, in several other scenarios, such as oscillon collisions and decay, the gravitational metric plays a passive role, and thus, this simplified framework and the codes mentioned previously can be safely employed.


\section{Caveats in the lattice approach}
\label{sec:caveats}

Here we will mostly follow and expand Appendix C of \cite{del-Corral:2025fzz}, and focus on the implementation of the lattice code $\mathcal{C}$\texttt{osmo}$\mathcal{L}$\texttt{attice} \cite{Figueroa:2021yhd,Figueroa:2020rrl,Figueroa:2023xmq,Baeza-Ballesteros:2025tme}, as it is built upon \texttt{LATTICEEASY} \cite{Felder:2000hq} and the subsequent ones \cite{Felder:2007kat,Frolov:2008hy,Sainio:2009hm,Sainio:2012mw,Easther:2010qz,Child:2013ria,Lozanov:2019jff,Huang:2011gf,Figueroa:2021yhd,Figueroa:2020rrl,Figueroa:2023xmq,Baeza-Ballesteros:2025tme}. This code considers that the only inhomogeneity is in the scalar field $\phi$. That is
\begin{equation}
    \phi(t,\bm x)=\bar\phi(t)+\delta\phi(t,\bm x),
\end{equation}
with $\bar\phi$ being the background solution and $\delta\phi$ the fluctuations around the homogeneous background. Here, the scalar field evolves as
\begin{equation}
\ddot{\phi}+3H\dot{\phi}-\frac{\nabla^{2}\phi}{a^{2}}+V_{,\phi}=0,
\label{eq:KG}
\end{equation}
where a dot denotes differentiation with respect to cosmic time $t$, $a$ is the scale factor, $V(\phi)$ the scalar field potential, and $H=\dot{a}/a$ the Hubble rate. The latter is determined from a volume average, represented by $\langle\dots\rangle_{\text{lat}}$, over the local stress-energy tensor
\begin{equation}
H^{2}=\frac{1}{3\Mpl^{2}}\,\bigg\langle\,\frac{\dot{\phi}^{2}}{2}+\frac{(\nabla\phi)^{2}}{2a^{2}}+V(\phi)\,\bigg\rangle_{\text{lat}} .
\label{eq:FriedmannAvg}
\end{equation}
As one can see, this prescription effectively replaces the local Einstein equations,
\begin{equation}
G_{\mu\nu}[g_{\mu\nu}(t,\bm x)] = 8\pi G T_{\mu\nu}[\phi(t,\bm x)],\label{eq:Einstein}
\end{equation}
with a single background equation sourced by the averaged energy density. While this provides a tractable description of the mean expansion, it does not capture the local metric fluctuations $\delta g_{\mu\nu}(t,\bm x)$ that are sourced by inhomogeneities in the scalar field $\delta\phi(t,\bm x)$. The lattice prescription removes those metric fluctuations by hand, which is a structural modification of General Relativity (GR). 

The lattice treatment can be understood as a mean-field approximation to the gravitational sector.  The scalar field is evolved with its full spatial dependence, so that fragmentation, gradients, and the formation of localized configurations are retained, while the geometry is represented only by a single homogeneous scale factor $a(t)$.  The effect of the inhomogeneous scalar field on the geometry is then reduced to its spatially averaged energy density, which determines the evolution of $a(t)$.  This approximation is reasonable as long as the local inhomogeneities of the scalar field do not produce an important local gravitational response.  The difficulty is that oscillon formation proceeds precisely through the growth and localization of those inhomogeneities. As the initially nearly homogeneous condensate fragments, the local energy density and gradient energy become increasingly different from their spatial averages. The lattice prescription nevertheless allows these local structures to evolve in a geometry that responds only to the averaged energy density.  In this sense, the approximation treats the matter sector locally but the gravitational sector only globally.

Whether this separation remains self-consistent during preheating is therefore not obvious and has to be checked from the Einstein equations themselves.  For a mode well inside the Hubble radius, the linearized Hamiltonian constraint reduces to the Poisson relation\footnote{In Newtonian gauge, the linearized $00$ Einstein equation for a canonical scalar field can be written as
\begin{equation}
    -\frac{k^2}{a^2}\Phi_k-3H\left(\dot{\Phi}_k+H\Phi_k\right)=\frac{\delta\rho_k}{2M_{\rm Pl}^2},
\end{equation}
up to the overall sign convention used for the Fourier transform and $\delta\rho_k$.  For sub-Hubble modes in the Poisson regime, $k/a\gg H$ and $|3H(\dot{\Phi}_k+H\Phi_k)|\ll (k^2/a^2)|\Phi_k|$, this reduces to $(k^2/a^2)\Phi_k\simeq-\delta\rho_k/(2M_{\rm Pl}^2)$.}
\begin{equation}
    \frac{k^2}{a^2}\Phi_k\simeq-\frac{\delta\rho_k}{2M_{\rm Pl}^2},
\end{equation}
and therefore, using $3M_{\rm Pl}^2H^2=\bar\rho$,
\begin{equation}
    |\Phi_k|\simeq\frac{3}{2}\left(\frac{aH}{k}\right)^2\left|\frac{\delta\rho_k}{\bar\rho}\right|.\label{Phieq}
\end{equation}
Thus even a sizeable density contrast can produce only a weak metric perturbation when its physical wavelength is sufficiently shorter than the Hubble scale. The fixed-FLRW approximation is therefore controlled only if the nonzero-wavenumber gravitational modes generated by the scalar inhomogeneities remain suppressed, more precisely if
\begin{equation}
    \left(\frac{aH}{k}\right)^2\left|\frac{\delta\rho_k}{\bar\rho}\right|\ll 1
\end{equation}
for the modes relevant to the dynamics.  The physical reasoning behind the fixed-FLRW approximation is therefore well motivated in an appropriate regime.  Oscillons are localized on scales that can be much shorter than the Hubble radius, and for $k/a\gg H$ the Hamiltonian constraint suppresses the metric potential by a factor $(aH/k)^2$ relative to the density contrast. A large scalar-field or density inhomogeneity can consequently coexist with a weak gravitational potential.  This motivates treating the scalar sector locally and nonlinearly while retaining gravity only through the homogeneous expansion. The approximation is reliable, however, only if the metric-induced terms also give perturbatively small corrections to the dynamics of the modes being amplified. This condition is distinct from the smallness of $\Phi_k$ itself. During a resonant instability, a metric contribution that is small in absolute magnitude can enter at the same order as the periodic modulation that determines the amplification. It can therefore modify the instability band and its growth rate even while $|\Phi_k|\ll1$. The relevant question is thus not only whether the metric potential is small, but whether the metric contribution is dynamically subleading in the instability problem.


The instabilities relevant to this discussion arise already within linear perturbation theory. For a canonical inflaton, the coupled scalar--metric degree of freedom is described by the Mukhanov--Sasaki variable
\begin{equation}
    v_k=a\left(\delta\phi_k+\frac{\dot{\bar\phi}}{H}\Phi_k\right),\qquad z=\frac{a\dot{\bar\phi}}{H}.
    \label{eq:vk}
\end{equation}
Combining the perturbed Einstein and Klein--Gordon equations gives
\begin{equation}
    v_k''+\left(k^2-\frac{z''}{z}\right)v_k=0,
    \label{eq:MS-preheating}
\end{equation}
where primes denote derivatives with respect to conformal time, $d\eta=dt/a$. During preheating, the oscillations of $\bar\phi$ modulate the effective frequency of this equation. For suitable wavenumbers, the modulation reinforces the perturbation over successive oscillations, producing parametric amplification. The instability bands identify the modes for which this cumulative growth occurs.

This mechanism can be exhibited explicitly when the inflaton dominates the background and oscillates, for instance, near a quadratic minimum, $V(\bar\phi)\simeq m^2\bar\phi^2/2$. Once $H/m\ll1$, the background takes the approximate form $\bar\phi=\phi_{\rm amp}(t)\sin(mt)$, with $\phi_{\rm amp}\propto a^{-3/2}$. Defining $\widetilde v_k=a^{1/2}v_k$ and $\tau=mt+\pi/4$, Eq.~\eqref{eq:MS-preheating} reduces, at leading order in $H/m$, to the Mathieu equation \cite{Martin:2019nuw}
\begin{equation}
    \begin{aligned}
        \frac{d^2\widetilde v_k}{d\tau^2}+\left[A_k-2q\cos(2\tau)\right]\widetilde v_k&\simeq0,\\
        A_k=1+\frac{k^2}{a^2m^2},\qquad q&\simeq\frac{3H}{m}.
    \end{aligned}
    \label{eq:MS-Mathieu}
\end{equation}
The coefficients vary slowly over an inflaton oscillation. For fixed coefficients, Floquet theory gives solutions $\widetilde v_k=e^{\mu_k\tau}P_k(\tau)$, where $P_k$ is a periodic function and $\mu_k$ is the so-called Floquet exponent. In the first narrow resonance band, this exponent is approximately given by
\begin{equation}
    \mu_k\simeq\frac12\sqrt{q^2-(A_k-1)^2},\qquad|A_k-1|<q.
    \label{eq:MS-Floquet}
\end{equation}
Consequently, the resonant range is $0<k/a<\sqrt{3Hm}$, whose sub-Hubble part supports the growth of density perturbations~\cite{del-Corral:2023apl,Jedamzik:2010dq}.

Well inside this band, $\mu_k\simeq3H/(2m)$, and the growing envelope behaves as $\widetilde v_k\propto a^{3/2}$, or $v_k\propto a$. For the corresponding growing solution, averaging over the rapid inflaton oscillations gives an approximately constant Bardeen potential and a density contrast $\delta_k\equiv\delta\rho_k/\bar\rho\propto a$. Thus, the metric potential can remain small while participating in a growing density mode.

For anharmonic potentials, whose minimum is not quadratic, the oscillating potential curvature $V_{,\phi\phi}(\bar\phi)$ supplies additional modulation, leading to a more general Hill equation and potentially much faster self-resonance~\cite{del-Corral:2024vcm}. In the quadratic example above, $V_{,\phi\phi}=m^2$ is constant: the leading resonant modulation instead originates from the scalar--metric coupling contained in Eq.~\eqref{eq:MS-preheating}. Although $q\ll1$, this gravitational contribution determines the resonance band and its growth rate. Its dynamical importance therefore requires examining the coupled equations for the relevant modes, as we do below.

\subsection{The momentum constraint}  

We should write the remaining Einstein equations in addition to the result of $00$ component, Eq.~\eqref{Phieq}, to verify whether the metric fluctuations are of the same order as the scalar-field fluctuations. Since we deal with a scalar field, there is no anisotropic stress, and the two Bardeen potentials are equal, $\Phi=\Psi$. Then, the momentum constraint reads, exactly within linearized GR and in Fourier space \cite{Mukhanov1992,Baumann0907}:
\begin{equation}
    \dot\Phi_k+H\Phi_k=\frac{1}{2\Mpl^{2}}\,\dot{\bar\phi}\,\delta\phi_k.
    \label{eq:Momentumconstraint}
\end{equation}
Dividing by $H$ and using the definition of the first slow-roll parameter
\begin{equation}
    \eps_H=-\frac{\dot{H}}{H^2}=\frac12\left(\frac{\dot{\bar\phi}}{H\Mpl}\right)^2,
\end{equation}
we obtain
\begin{equation}
    \frac{\dot\Phi_k}{H}+\Phi_k=\sqrt{2\eps_H}\;\frac{\delta\phi_k}{2\Mpl}.
    \label{eq:MomentumconstraintRescaled}
\end{equation}
This equation is an identity of linearized GR, and therefore does not depend on the inflationary model or on the gauge choice, once $\Phi$ has been defined. It presents two regimes:

\begin{itemize}

    \item \textit{Slow-roll inflation}. Here, $\eps_H\ll 1$, and the right-hand side is suppressed by the smallness of $\eps_H$. Thus, in this regime, one can safely ignore the metric fluctuations.

    \item \textit{Preheating}. The inflaton field oscillates around the minimum of its potential, which makes $\eps_H$ oscillate between $0$ and $3$ with an order-unity time-average over an oscillation period. Here, the hierarchy that justified ignoring $\Phi_k$ during inflation no longer applies.

\end{itemize}

Inside the instability band \cite{del-Corral:2023apl} or, at least, after the strong self-resonance phase \cite{del-Corral:2024vcm} (if it exists), the metric perturbation oscillates with constant amplitude. One can show that $|\dot\Phi_k/H|\ll|\Phi_k|$ \cite{del-Corral:2023apl}. Then, Eq.~\eqref{eq:Momentumconstraint} further reduces to
\begin{equation}
\Phi_k\simeq \frac{\delta\phi_k}{\Mpl},
\label{eq:Phisimdphi}
\end{equation}
so that the Bardeen potential and the field perturbation are of the same parametric size, fixed entirely by the background. Therefore, the statement that metric perturbations are small during resonance \cite{Amin:2011hj} is shown not to be precise, and we believe that the misunderstanding comes from extending the slow-roll behavior of the momentum constraint \eqref{eq:MomentumconstraintRescaled} to a non-slow-roll phase.


\subsection{The perturbed Klein--Gordon equation}
\label{subsec:perturbed-KG}

The dynamical significance of the metric response can be examined directly in the perturbed Klein--Gordon equation. For a canonical scalar field, in Newtonian gauge and Fourier space, this equation reads
\begin{equation}
    \begin{aligned}
        \ddot{\widetilde{\delta\phi}}_k&+\left[\frac{k^2}{a^2}+V_{,\phi\phi}-\frac94H^2-\frac32\dot H\right]\widetilde{\delta\phi}_k
        \\&=a^{3/2}\left(4\dot{\bar\phi}\dot\Phi_k-2V_{,\phi}\Phi_k\right)\ ,
    \end{aligned}
    \label{eq:dphi}
\end{equation}
where $\widetilde{\delta\phi}_k \equiv a^{3/2}\delta\phi_k$ and derivatives of the potential are evaluated on $\bar\phi(t)$. The lattice codes \cite{Felder:2000hq,Felder:2007kat,Frolov:2008hy,Sainio:2009hm,Sainio:2012mw,Easther:2010qz,Child:2013ria,Lozanov:2019jff,Easther:2006vd,Easther:2007vj,Huang:2011gf,Figueroa:2021yhd,Figueroa:2020rrl,Figueroa:2023xmq,Baeza-Ballesteros:2025tme} do not consider this equation and simply evolve the background field equation \eqref{eq:KG} assuming an inhomogeneous scalar field, but still neglecting metric fluctuations, which can be seen as the equivalent of ignoring the right-hand side of Eq.~\eqref{eq:dphi}. 

The momentum constraint \eqref{eq:Momentumconstraint} fixes the derivative of the potential appearing in these terms. Substituting it into Eq.~\eqref{eq:dphi} gives the exact identity
\begin{equation}
    \begin{aligned}
        4\dot{\bar\phi}\dot{\Phi}_k-2V_{,\phi}\Phi_k={}\frac{2\dot{\bar\phi}^{\,2}}{M_{\rm Pl}^2}\delta\phi_k-2\left(V_{,\phi}+2H\dot{\bar\phi}\right)\Phi_k.
    \end{aligned}
    \label{eq:KG-constrained-source}
\end{equation}
This displays explicitly how the gravitational source depends on both the background motion and the field fluctuation. Its importance requires evaluating these couplings for the modes being amplified. A bound on $|\Phi_k|$ obtained from the momentum constraint alone does not establish their effect on the growth rate.

The contribution of the constraints with metric fluctuations is taken into account in the equation for the rescaled Mukhanov--Sasaki variable $\widetilde v_k$ introduced above. Eliminating the constrained metric perturbations gives \cite{del-Corral:2023apl}
\begin{equation}
    \begin{aligned}
        \ddot{\widetilde v}_k+\Bigg[\frac{k^2}{a^2}+V_{,\phi\phi}+\Delta m_{\rm grav}^2-\frac94H^2-\frac32\dot H\Bigg]\widetilde v_k=0,
    \end{aligned}
\label{eq:MS-gravitational-mass}
\end{equation}
where
\begin{equation}
    \Delta m_{\rm grav}^2=-\frac{1}{M_{\rm Pl}^2a^3}\frac{d}{dt}\left(\frac{a^3\dot{\bar\phi}^{\,2}}{H}\right).
    \label{eq:gravitational-mass}
\end{equation}
The derivative term is the gravitational coupling generated by enforcing the Einstein constraints. We can notice that imposing $\Phi_k\to 0$ in \eqref{eq:MS-gravitational-mass} does not reduce to \eqref{eq:dphi}, which is because the momentum constraint \eqref{eq:Momentumconstraint} implies that switching off the metric potential automatically gives vanishing scalar field fluctuations. This demonstrates the prominence of the Mukhanov-Sasaki variable in deducing both the gravitational and matter field dynamics. 

The role of \eqref{eq:gravitational-mass} is particularly transparent in the quadratic regime discussed above. Using $\bar\phi\simeq\phi_{\rm amp}(t)\sin(mt)$, $\phi_{\rm amp}\propto a^{-3/2}$, and $m^2\phi_{\rm amp}^2\simeq6M_{\rm Pl}^2H^2$, one finds
\begin{equation}
    \Delta m_{\rm grav}^2=6mH\sin(2mt)+\mathcal{O}(H^2).
    \label{eq:gravitational-resonant-term}
\end{equation}
With the phase convention $\tau=mt+\pi/4$, this becomes $-6mH\cos(2\tau)$ and supplies the Mathieu parameter $q\simeq3H/m$. The constant potential curvature $V_{,\phi\phi}=m^2$ sets the oscillation frequency, while the gravitational contribution provides the leading modulation responsible for this resonance.

Although the modulation is suppressed by $H/m$ relative to $m^2$, its cumulative effect acts over the expansion time: well inside the quadratic instability band, $m\mu_k\simeq3H/2$. Consequently, omitting the gravitational coupling removes the leading mechanism responsible for this growing mode. For anharmonic potentials, the time dependence of $V_{,\phi\phi}(\bar\phi)$ supplies additional modulation, whose relative importance must be assessed together with Eq.~\eqref{eq:gravitational-mass}. The appropriate comparison is therefore between the instability bands and growth rates of the coupled system and those obtained from the FLRW prescription.

\subsection{Spatial averaging within the ADM formalism}

The standard claim to defend Eq.~\eqref{eq:FriedmannAvg} is that, although it is not the local Einstein equations, it captures the average effect of inhomogeneities, with the metric fluctuations reabsorbed into a spatially-averaged Hubble rate. In what follows, we show that this argument is closed by a proper general-relativistic averaging of the evolution equations. In essence, we show that the proper-volume average of the local ADM Hamiltonian constraint over a spatial domain $\mathcal{D}$ does not reduce to the Friedmann equation employed in the fixed-FLRW lattice prescription. The reason is that averaging and nonlinear gravitational evolution do not commute. To show this, we follow the standard ADM (Arnowitt-Deser-Misner) 3+1 decomposition \cite{Arnowitt:1962hi} of the metric tensor into spatial hypersurfaces that evolve in time. The spacetime metric in ADM form is given by
\begin{equation}\label{eq:adm}
    ds^2=-N^2dt^2+h_{ij}\left(dx^i+N^i dt\right)\left(dx^j+N^j dt\right),
\end{equation}
where $N$, $N^i$, and $h_{ij}$ are the lapse, shift, and induced spatial metric. We use the expansion-tensor convention
\begin{equation}\label{eq:theta-def}
    \Theta_{ij}=\frac{1}{2N}\left(\dot h_{ij}-D_iN_j-D_jN_i\right),\quad\theta=h^{ij}\Theta_{ij},
\end{equation}
so that $\Theta_{ij}=-K_{ij}$, relative to the common numerical-relativity sign convention, and with $K_{ij}$ being the extrinsic curvature tensor. We can decompose the expansion tensor into its trace and traceless parts as follows
\begin{equation}\label{eq:shear-def}
    \Theta_{ij}=\frac{1}{3}\theta h_{ij}+\sigma_{ij},\qquad\sigma^2\equiv\frac{1}{2}\sigma_{ij}\sigma^{ij},
\end{equation}
where $\sigma_{ij}$ is the shear tensor and $\sigma^2$ is the shear scalar. With $8\pi G=\mpl^{-2}$, the local Hamiltonian and the local momentum constraints are thus given by
\begin{align}
    \Rthree+\frac{2}{3}\theta^2-2\sigma^2&=\frac{2}{\mpl^2}\rho_n,\label{eq:local-H}
    \\
    D_j\!\left(\Theta^j{}_i-\delta^j{}_i\theta\right)&=\frac{1}{\mpl^2}\mathcal J_i.\label{eq:local-M}
\end{align}
Here, $\rho_n$ and $\mathcal{J}_i$ are the normal-normal and normal-spatial projections of Einstein's equations \cite{Gourgoulhon:2007ue,Baumgarte:2010ndz}, defined as
\begin{equation}\label{eq:rho-J-def}
    \rho_n\equiv T_{\mu\nu}n^\mu n^\nu,\qquad\mathcal J_i\equiv h_i{}^\mu n^\nu T_{\mu\nu},
\end{equation}
where $n^\mu$ is the future-directed unit normal to the slice. For a canonical scalar field, these quantities are given by
\begin{align}
    u_{\mu}&=\frac{\partial_{\mu}\phi}{\sqrt{-g^{\mu\nu}\partial_{\mu}\phi\partial_{\nu}\phi}}\\
    \rho_n&=\frac{1}{2N^2}\left(\dot\phi-N^i\partial_i\phi\right)^2+\frac{1}{2}h^{ij}\partial_i\phi\partial_j\phi+V(\phi),\label{eq:rho-n}
    \\
    \mathcal J_i&=\frac{1}{N}\left(\dot\phi-N^j\partial_j\phi\right)\partial_i\phi,\label{eq:J-i}
\end{align}
which shows that both constraints are sourced locally by the same inhomogeneous field that the lattice evolves. We now take the proper-volume average of Eq.~\eqref{eq:local-H} over a compact spatial domain $\mathcal D$. Since this is an instantaneous average of the local Hamiltonian constraint on a single hypersurface, no separate averaged evolution formalism is required. We define the proper volume and the proper-volume average on a single time slice by
\begin{equation}\label{eq:proper-average}
    \VD\equiv\int_{\mathcal D}d^3x\,\sqrt h,\qquad\avgV{F}\equiv\frac{1}{\VD}\int_{\mathcal D}d^3x\,\sqrt h\,F.
\end{equation}
Applying this linear integration operation to \eqref{eq:local-H} gives the exact instantaneous identity
\begin{equation}\label{eq:averaged-H-raw}
    \avgV{\Rthree}+\frac{2}{3}\avgV{\theta^2}-2\avgV{\sigma^2}=\frac{2}{\mpl^2}\avgV{\rho_n}.
\end{equation}
Define the mean normal-expansion rate and its fluctuation by
\begin{equation}\label{eq:Htheta-def}
    H_\theta\equiv\frac{1}{3}\avgV{\theta},\qquad\delta\theta\equiv\theta-\avgV{\theta},
\end{equation}
so that $\avgV{\theta^2}$ can be expressed as
\begin{equation}\label{eq:theta-square}
    \avgV{\theta^2}=9H_\theta^2+\avgV{(\delta\theta)^2}.
\end{equation}
Substituting this into \eqref{eq:averaged-H-raw} yields
\begin{equation}\label{eq:averaged-H-exact}
    3H_\theta^2=\frac{1}{\mpl^2}\avgV{\rho_n}-\frac{1}{2}\avgV{\Rthree}-\frac{1}{3}\avgV{(\delta\theta)^2}+\avgV{\sigma^2}.
\end{equation}
This equation is nothing more than the spatial integral of the local ADM constraint. In particular, it does not use the Buchert evolution equations \cite{Buchert2000,Buchert2002}, a dust description, or a commutation rule between averaging and time evolution. As shown in Eq.~\eqref{eq:FriedmannAvg}, the scalar energy density inserted into the lattice codes is
\begin{equation}\label{eq:rho-lat}
    \rho_{\rm lat}\equiv\frac{1}{2}\dot\phi^{\,2}+\frac{(\grad\phi)^2}{2a^2}+V(\phi),
\end{equation}
so that the Friedmann equation they use is given by
\begin{equation}\label{eq:lattice-Friedmann}
    3H_{\rm lat}^2=\frac{1}{\mpl^2}\avglat{\rho_{\rm lat}},
\end{equation}
which is then used to evolve or check the homogeneous scale factor. The comparison with GR is now immediate.  Subtract Eq.~\eqref{eq:lattice-Friedmann} from the directly averaged Hamiltonian constraint Eq.~\eqref{eq:averaged-H-exact}. This gives the single exact identity
\begin{equation}\label{eq:master-comparison}
    \begin{aligned}3\left(H_\theta^2-H_{\rm lat}^2\right)={}&\frac{1}{\mpl^2}\left(\avgV{\rho_n}-\avglat{\rho_{\rm lat}}\right)-\frac{1}{2}\avgV{\Rthree}
    \\
    &-\frac{1}{3}\operatorname{Var}_V(\theta)+\avgV{\sigma^2},\end{aligned}
\end{equation}
where we have defined the volume-averaged variance of $\theta$ as
\begin{equation}
    \operatorname{Var}_V(\theta)=\avgV{\theta^2}-\avgV{\theta}^{\,2}=\avgV{(\delta\theta)^2}.
\end{equation}
The terms in Eq.~\eqref{eq:master-comparison} can now be evaluated explicitly using the Newtonian-gauge scalar metric
\begin{equation}\label{eq:newtonian-metric}
    ds^2=-(1+2\Phi)dt^2+a^2(t)(1-2\Phi)\delta_{ij}dx^i dx^j.
\end{equation}
This is equivalent to choosing the following lapse, shift, and induced spatial metric in the ADM decomposition
\begin{equation}
    N=\sqrt{1+2\Phi},\qquad N^i=0,\qquad h_{ij}=a^2(1-2\Phi)\delta_{ij}.
\end{equation}
We work in the regular domain $1+2\Phi>0$ and $1-2\Phi>0$, so that the lapse is real and the spatial metric is positive definite. Within this conformally flat choice, the shear $\sigma_{ij}$ vanishes and the normal expansion is given by
\begin{equation}
    \theta=\frac{3}{\sqrt{1+2\Phi}}\left(H-\frac{\dot\Phi}{1-2\Phi}\right),
\end{equation}
which shows the spatial dependence explicitly through the metric fluctuation $\Phi=\Phi(t,\bm{x})$. Here, $H=\tfrac{\dot{a}}{a}$ is the standard Hubble rate. The exact intrinsic scalar curvature of a constant-$t$ slice is
\begin{equation}\label{eq:exact-R}
    \Rthree =\frac{4}{a^2}\frac{\nabla^2\Phi}{(1-2\Phi)^2}+\frac{6}{a^2}\frac{(\grad\Phi)^2}{(1-2\Phi)^3}.
\end{equation}
This follows directly from the conformal transformation law for the three-curvature (see Sec.~6.3.3 of Ref.~\cite{Gourgoulhon:2007ue}). The normal-frame scalar energy is now
\begin{equation}\label{eq:rho-n-newtonian}
    \rho_n=\frac{\dot\phi^{\,2}}{2(1+2\Phi)}+\frac{(\grad\phi)^2}{2a^2(1-2\Phi)}+V(\phi),
\end{equation}
and the local Hamiltonian constraint becomes
\begin{equation}\label{eq:exact-local-newtonian}
   \frac{1}{3} \theta^2+\frac{1}{2}\Rthree=\frac{1}{\mpl^2}\rho_n.
\end{equation}
Finally, the metric correction to the energy density used in the lattice approach is explicitly
\begin{equation}\label{eq:matter-metric-correction}
    \rho_n-\rho_{\rm lat}=-\frac{\Phi}{1+2\Phi}\dot\phi^{\,2}+\frac{\Phi}{a^2(1-2\Phi)}(\grad\phi)^2.
\end{equation}
Hence, even the local matter expression retained by the lattice codes is not the normal-frame energy in the Hamiltonian constraint when $\Phi\neq0$.  Moreover, as shown in \cite{Buchert2000}, even if the initial lattice data considers a vanishing average curvature, it will be generated in the course of structure formation by the actual inhomogeneities that the resonance amplifies. This further shows that the lattice prescription implements neither the local Einstein equation~\eqref{eq:Einstein} nor the averaged equation~\eqref{eq:averaged-H-exact}. It implements an approximation in which nonlinear matter dynamics are retained, while gravitational effects, at the local and averaged level, are neglected. The claim that neglecting the metric fluctuations is innocuous because the relevant physics has been absorbed into $H$ is incompatible with a proper ADM-based averaging. Therefore, the identification of $H_{\text{lat}}$ in Eq.~\eqref{eq:lattice-Friedmann} with the averaged Hubble rate $H_{\theta}$ is not correct.


\subsection{Numerical illustration}

\begin{figure}[t]
    \centering
    \includegraphics[width=\linewidth]{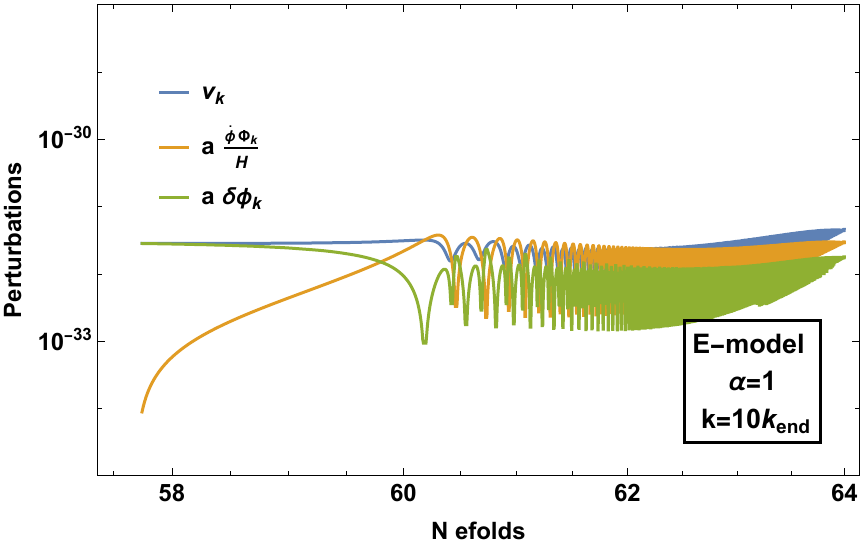}
    \caption{Components of the Mukhanov--Sasaki variable, Eq.~\eqref{eq:vk}. Before the end of inflation, $N\simeq 60$, the metric contribution (orange curve) is parametrically suppressed by $\sqrt{2\eps_H}\ll 1$ relative to the matter contribution (green curve). Ignoring $\Phi_k$ in this regime is a controlled approximation. After the end of inflation and the beginning of the preheating phase, $N\simeq 60$, $\eps_H$ becomes of order unity, and the orange curve climbs into the same range as the green, and the suppression that justified ignoring $\Phi_k$ disappears. Lattice codes that solve the Klein-Gordon Eq.~\eqref{eq:KG} on the averaged background~\eqref{eq:FriedmannAvg} retain just the green contribution and discard the orange one. The underlying model is the Starobinsky model,  equivalent to an E-model with $\alpha=1$.}
    \label{fig:staro}
\end{figure}

\begin{figure*}
    \includegraphics[width=\textwidth]{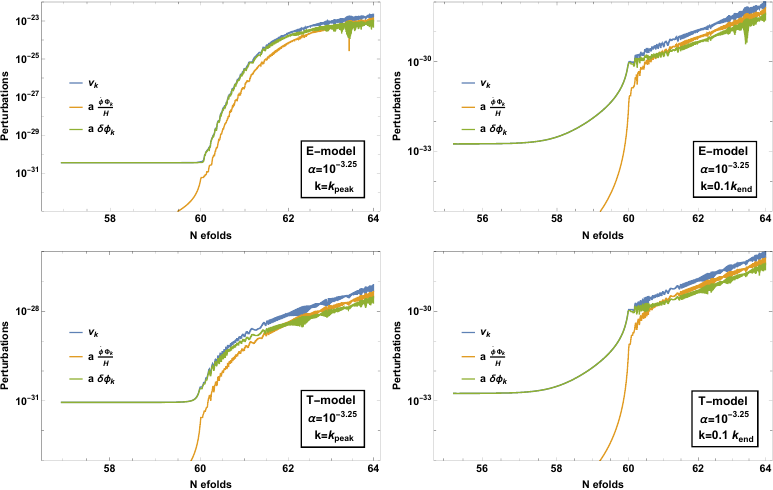}
    \caption{The same as Fig.~\ref{fig:staro} but for different $\alpha$-attractor models. See the insets on each plot for the details about the specific model and value of $\alpha$.}
    \label{fig:components}
\end{figure*}

To show the importance and validity of the specific claim we make in this work, we show the evolution of the Mukhanov-Sasaki variable for a case of moderate instabilities, the Einstein frame Starobinsky model \cite{Starobinsky:1980te,Whitt:1984pd,Maeda:1988ab}, in Fig.~\ref{fig:staro}. The figure displays the two components of Eq.~\eqref{eq:vk}, the metric contribution in orange, the scalar field contribution in green, and the full Mukhanov-Sasaki variable in blue. For completeness, we also show in Fig.~\ref{fig:components} the same plot for the $\alpha$-attractors \cite{Kallosh:2013daa,Kallosh:2013hoa,Kaiser:2013sna,Kallosh:2015lwa} T- and E-models, where the strong self-resonance instabilities are present \cite{del-Corral:2024vcm}. The figure is taken from \cite{del-Corral:2025fzz}, and the wavenumbers chosen in this case are the corresponding ones to the peak of amplification due to instabilities for each value of the parameter $\alpha$. For the Starobinsky model, where the strong instabilities are absent, the metric fluctuations are still of the same order as the scalar field fluctuations after the end of inflation. In both figures, one can see that before the end of inflation ($N\lesssim 60$), the metric contribution is two or more orders of magnitude smaller than the scalar field piece. This is the regime in which $\sqrt{2\eps_H}\ll 1$ holds. The metric piece climbs immediately at the end of inflation and tracks the matter piece within a factor of $\mathcal{O}(1)$ for the entire amplification phase, in exact agreement with Eq.~\eqref{eq:Phisimdphi}. 


\section{Numerical Relativity as the Consistent Framework}
\label{sec:numerical-relativity}

In this section, we propose the well-known, consistent numerical and non-linear framework that considers both metric and scalar field fluctuations: The BSSN formalism. However, we first give some context on the state of the art in this field and how much it differs from and improves the ADM decomposition shown in the previous section. 

The pillars of modern cosmology are built on the perturbed Einstein equations, in which the metric perturbations are the main quantities employed. Some examples are the CMB power spectrum through the Sachs-Wolfe and integrated Sachs-Wolfe effects, the late-time integrated Sachs-Wolfe signature of dark energy, the CMB lensing, the weak gravitational lensing of large-scale structure, the linear matter power spectrum used in baryon acoustic oscillation analyses, and the Newtonian potential responsible for nonlinear structure formation. The linear perturbation theory of Bardeen, Kodama, and Sasaki, and Mukhanov, Feldman, and Brandenberger~\cite{Mukhanov1992,Baumann0907} is the framework within which all of these signals are computed.

The numerical relativity programme of the last few years implements the full ADM decomposition \cite{Arnowitt:1962hi}. However, it has some caveats. When one applies this splitting to the Einstein equations, one arrives at evolution equations for $h_{ij}$ and $K_{ij}$ and the Hamiltonian and momentum constraints, which must hold on every hypersurface. However, this set of Einstein equations is weakly hyperbolic since \cite{Baumgarte:1998te,Baumgarte:2010ndz,Baumgarte:2021skc} some of the terms containing higher-order derivatives cannot be separated into a full set of independent modes. This implies that small changes in the initial data do not produce small changes in the solution for a finite time, which makes the ADM decomposition numerically unstable.

The BSSN (Baumgarte–Shapiro–Shibata–Nakamura) formalism \cite{Nakamura:1987zz,Shibata:1995,Baumgarte:1998te,Baumgarte:2010ndz,Baumgarte:2021skc} addresses this instability and improves the ADM decomposition by rewriting the dynamical variables: the spatial metric is split into a conformal metric $\tilde h_{ij}$ and a volume factor $\gamma$, and the extrinsic curvature into the trace and trace-free parts, similar to the Buchert procedure \cite{Buchert2000,Buchert2002}. Also, it promotes the conformal connection functions, $\tilde\Gamma^i=\tilde h^{jk}\tilde\Gamma^i_{jk}$, to be independent variables as well. The $\tilde\Gamma^i_{jk}$ are the affine connection coefficients computed with the conformal metric $\tilde h_{ij}$. These changes make some of the terms in the Einstein equations containing problematic second-derivative combinations be separated into first-order variables, and the constraints are used to simplify the problematic higher-derivative terms. In fact, with a suitable choice of lapse and shift, the highest-derivative terms of the equations have a complete set of characteristic modes that allows for proper control of the errors. It is important to remark that the BSSN formalism is just a reformulation of the Einstein equations, and one needs to further choose, for instance, gauge conditions for the lapse and shift, spatial discretization methods, or a time integrator. In general, when we refer to the BSSN formalism, we refer to the combination of the BSSN equations with a suitable numerical scheme. Several other conformal formulations of the Einstein equations exist. For instance, the General Harmonic (GH) formulation \cite{Pretorius:2004jg}, or the formalisms based on the Z4 formulation \cite{Bona:2003fj}, such as the Conformal-Covariant-Z4 (CCZ4) \cite{Alic:2011gg}, or the Conformal-Z4 (Z4c) \cite{Weyhausen:2011cg}. In this work, we focus on the BSSN for its historical importance. 

Through the BSSN formalism, or in general any formalism based on numerical relativity, the evolution of the metric alongside the matter sector is, in our opinion, the correct framework. Particularly, the \texttt{GRChombo} collaboration~\cite{GRChombo} has applied this framework to the preheating context in~\cite{Aurrekoetxea2023,deJong2023}, with initial data constructed using the CTTK method~\cite{Aurrekoetxea2022CTTK}, which can be seen as a complement to the BSSN. The former prepares constraint-satisfying data at the initial time, while the BSSN then evolves those data forward in time. Another example is the \texttt{GABERel} code \cite{Giblin:2019nuv}, an adaptation of \texttt{GABE} \cite{Child:2013ria} that in this case does include the metric fluctuations and employs the BSSN formalism \cite{Grutkoski:2025ygs,Adshead:2023mvt,Baumgarte:2026igz}. In general, these simulations evolve $\Phi_k$ (in the appropriate ADM language) as a dynamical variable, where the Hamiltonian and momentum constraints are checked at every time step for consistency. Although we do not endorse every quantitative conclusion of \cite{Aurrekoetxea2023,deJong2023,Aurrekoetxea2022CTTK,Grutkoski:2025ygs,Adshead:2023mvt,Giblin:2019nuv}, we use these works as an example of the proper inclusion of metric fluctuations in numerical relativity, and therefore, the main criticism of this work does not apply in that case.

Finally, several quantities commonly extracted from fixed-FLRW lattice simulations are directly sensitive to the gravitational sector omitted in this approximation. These include: i) the thresholds for oscillon formation, which are determined from the nonlinear growth and localisation of the scalar inhomogeneities; ii) the subsequent evolution of the effective equation of state, including the prolonged regime in which \(\langle w_\phi\rangle\simeq0\) is associated with an oscillon-dominated phase; iii) criteria for fragmentation or destruction of the homogeneous condensate based on the growth of scalar-field fluctuations relative to the homogeneous mode; and iv) the gravitational-wave signal sourced during the nonlinear evolution. In the latter case, the second-order tensor source contains contributions from both the scalar-field and scalar-metric sectors, including terms schematically of the form \(\partial\delta\phi\,\partial\delta\phi\) and \(\partial\Phi\,\partial\Phi\). Fixed-FLRW lattice calculations retain the former while omitting the scalar-metric contribution \cite{Antusch:2016con}. A constraint-consistent treatment \cite{del-Corral:2025fzz} is therefore required to determine how each of these predictions is modified once the gravitational response is included.

The analysis above points to two systematic frameworks for post-inflationary dynamics. In the nonlinear regime, numerical relativity evolves the scalar field and spacetime geometry simultaneously while enforcing the Einstein constraints, for example through the BSSN formulation \cite{Aurrekoetxea2023}. In the perturbative regime, while the curvature perturbation remains subunity, the coupled scalar--metric dynamics are captured by the Mukhanov--Sasaki equation \cite{del-Corral:2025fzz,del-Corral:2025lcp}. These approaches provide a consistent framework for determining the formation and evolution of post-inflationary structures, including oscillons.


\section{Conclusions}
\label{sec:conclusions}

In this work, we show that the effect of the metric perturbations on the inflaton field cannot be neglected during resonance, as it contradicts the linearized momentum constraint of GR, Eq.~\eqref{eq:Momentumconstraint}. The suppression of $\Phi_k$ during slow-roll inflation is guaranteed by the smallness of the first slow-roll parameter $\eps_H$, as shown in Eq.~\eqref{eq:MomentumconstraintRescaled}. However, the latter becomes of order unity at the onset of preheating, and the suppression no longer applies. We also show that the perturbed Klein-Gordon equation \eqref{eq:dphi} contains $\Phi_k$-dependent operators that are of the same order as the resonance mass term $V_{,\bar\phi\bar\phi}$, which should not be neglected. In fact, the Mukhanov-Sasaki variable $v_k$, the unique gauge-invariant scalar degree of freedom of the coupled metric-matter system, receives comparable contributions, in order of magnitude, from both the metric and the scalar field sectors.

Following the ADM-based averaging procedure, we also show that the spatially averaged Friedmann equation that the lattice codes integrate, Eq.~\eqref{eq:FriedmannAvg} or \eqref{eq:lattice-Friedmann}, even if the metric fluctuations are small, is neither the local Hamiltonian constraint nor a consistent general-relativistic average of it, as shown in Eq.~\eqref{eq:averaged-H-exact}. This is because, although it retains the average density part $\langle\rho\rangle$, it neglects the average scalar curvature $\langle^{(3)}\mathcal{R}\rangle_{V}$, the variance of the normal-expansion rate $\text{Var}_V(\theta)$, and the averaged shear $\langle\sigma^2\rangle_V$. In fact, we show that these quantities are themselves sourced not only by the perturbations of the scalar field that the lattice codes keep, which again signals inconsistencies in their approach, but also by the metric fluctuations that these codes ignore. 

This paper proposes a structural correction in the lattice approach towards a more general relativistic framework. The path forward, on the nonlinear side, is a genuine numerical relativity reformulation based on the ADM decomposition of the metric \cite{Arnowitt:1962hi} that can be achieved through numerous formalisms. Among them, the BSSN formalism \cite{Shibata:1995,Baumgarte:1998te,Nakamura:1987zz} builds on the improved ADM decomposition and in essence constitutes a stable version of it. On the linear side, we propose the Mukhanov-Sasaki equation, based on the linear cosmological perturbation theory, which has been implemented in multiple works \cite{del-Corral:2023apl,del-Corral:2024vcm,Martin:2019nuw,Martin:2020fgl,Jedamzik:2010dq,Ballesteros:2024eee,del-Corral:2025fca,del-Corral:2025lcp}. 

As a summary, the volume-averaged Friedmann prescription of the lattice approaches, and the predictions built on it, should be treated as a first approach to a much richer and more complex scenario, where the metric fluctuations play an important role in some scenarios of parametric instabilities. We intend to compare in future work the results of the lattice simulations with a full general relativistic numerical computation in particular scenarios where the metric perturbations amplify.


\begin{acknowledgments}

D.dC.\ acknowledges the support of the grant No. UMO-2021/42/E/ST9/00260 from the National Science Centre, Poland. K.S.K. acknowledges the support of the Royal Society Newton International Fellowship. This research was also funded by Funda\c{c}\~ao para a Ci\^encia e a Tecnologia grant number UIDB/MAT/00212/2020 and COST action 23130. The work of P.G. was supported in part by NSF Grant No. PHY-2412829 and by the World Research Hub (WRH) Program of the Institute of Science Tokyo. P.G. thanks Prof. Teruaki Suyama for his kind hospitality at the Institute of Science Tokyo.

\end{acknowledgments}

\bibliographystyle{apsrev4-2}
\bibliography{references}

@article{Bogolyubsky:1976yu,
    author = "Bogolyubsky, I. L. and Makhankov, V. G.",
    title = "{Lifetime of Pulsating Solitons in Some Classical Models}",
    journal = "Pisma Zh. Eksp. Teor. Fiz.",
    volume = "24",
    pages = "15--18",
    year = "1976"
}

@article{Bogolyubsky:1976pw,
    author = "Bogolyubsky, I. L.",
    title = "{Oscillating Particle-Like Solutions of Nonlinear Klein-Gordon Equation}",
    reportNumber = "JINR-E2-10129",
    journal = "JETP Lett.",
    volume = "24",
    pages = "535",
    year = "1976"
}

@article{Gleiser1994,
  author  = {Gleiser, M.},
  title   = {Pseudostable bubbles},
  journal = {Phys. Rev. D},
  volume  = {49},
  pages   = {2978},
  year    = {1994},
  eprint  = {hep-ph/9308279},
  archivePrefix = {arXiv}
}

@article{Copeland1995,
  author  = {Copeland, E. J. and Gleiser, M. and M\"uller, H.-R.},
  title   = {Oscillons: Resonant configurations during bubble collapse},
  journal = {Phys. Rev. D},
  volume  = {52},
  pages   = {1920},
  year    = {1995},
  eprint  = {hep-ph/9503217},
  archivePrefix = {arXiv}
}

@article{Amin:2010xe,
    author = "Amin, Mustafa A.",
    title = "{Inflaton fragmentation: Emergence of pseudo-stable inflaton lumps (oscillons) after inflation}",
    eprint = "1006.3075",
    archivePrefix = "arXiv",
    primaryClass = "astro-ph.CO",
    month = "6",
    year = "2010",
    journal = {arXiv preprint}
}

@article{Amin:2010dc,
    author = "Amin, Mustafa A. and Easther, Richard and Finkel, Hal",
    title = "{Inflaton Fragmentation and Oscillon Formation in Three Dimensions}",
    eprint = "1009.2505",
    archivePrefix = "arXiv",
    primaryClass = "astro-ph.CO",
    doi = "10.1088/1475-7516/2010/12/001",
    journal = "JCAP",
    volume = "2010",
    number = "12",
    pages = "001",
    year = "2010"
}

@article{LatticeEASY,
  author  = {Felder, G. N. and Tkachev, I.},
  title   = {{LATTICEEASY}: A program for lattice simulations of scalar fields in an expanding universe},
  journal = {Comput. Phys. Commun.},
  volume  = {178},
  pages   = {929},
  year    = {2008},
  eprint  = {hep-ph/0011159},
  archivePrefix = {arXiv}
}

@article{DEFROST,

  author = {Frolov, A. V.},
  title = {{DEFROST}: A New Code for Simulating Preheating after Inflation},
  journal = {JCAP},
  volume = "2008",
  number = "11",
  pages = "009",
  year = {2008},
  eprint = {0809.4904},
  archivePrefix = {arXiv}
}

@article{CUDAEasy,
  author  = {Sainio, J.},
  title   = {{CUDAEASY} - a {GPU} accelerated cosmological lattice program},
  journal = {Comput. Phys. Commun.},
  volume  = {181},
  pages   = {906},
  year    = {2010},
  eprint  = {0911.5692},
  archivePrefix = {arXiv}
}

@article{HLattice,
  author  = {Huang, Z.},
  title   = {The Art of Lattice and Gravity Waves from Preheating},
  journal = {Phys. Rev. D},
  volume  = {83},
  pages   = {123509},
  year    = {2011},
  eprint  = {1102.0227},
  archivePrefix = {arXiv}
}

@article{GABE,
  author  = {Child, H. L. and Giblin, Jr., J. T. and Ribeiro, R. H. and Seery, D.},
  title   = {Preheating with Non-Minimal Kinetic Terms},
  journal = {Phys. Rev. Lett.},
  volume  = {111},
  pages   = {051301},
  year    = {2013},
  eprint  = {1305.0561},
  archivePrefix = {arXiv}
}

@article{PSpectRe,

  author = {Easther, R. and Finkel, H. and Roth, N.},
  title = {{PSpectRe}: A Pseudo-Spectral Code for (P)reheating},
  journal = {JCAP},
  volume = "2010",
  number = "10",
  pages = "025",
  year = {2010},
  eprint = {1005.1921},
  archivePrefix = {arXiv}
}

@article{Mukhanov1992,
  author  = {Mukhanov, V. F. and Feldman, H. A. and Brandenberger, R. H.},
  title   = {Theory of cosmological perturbations},
  journal = {Phys. Rept.},
  volume  = {215},
  pages   = {203},
  year    = {1992}
}

@article{Baumann0907,
  author  = {Baumann, D.},
  title   = {{TASI} Lectures on Inflation},
  journal = {arXiv preprint},
  year    = {2009},
  eprint  = {0907.5424},
  archivePrefix = {arXiv}
}

@article{Buchert2000,
  author  = {Buchert, T.},
  title   = {On average properties of inhomogeneous fluids in general relativity. {I}. {D}ust cosmologies},
  journal = {Gen. Rel. Grav.},
  volume  = {32},
  pages   = {105},
  year    = {2000},
  eprint  = {gr-qc/9906015},
  archivePrefix = {arXiv}
}

@article{Buchert2002,
  author  = {Buchert, T.},
  title   = {On average properties of inhomogeneous fluids in general relativity. {II}. {P}erfect fluid cosmologies},
  journal = {Gen. Rel. Grav.},
  volume  = {33},
  pages   = {1381},
  year    = {2001},
  eprint  = {gr-qc/0102049},
  archivePrefix = {arXiv}
}

@article{GRChombo,
  author  = {Andrade, T. and Salo, L. Areste and Aurrekoetxea, J. C. and Bamber, J. and Clough, K. and Croft, R. and de Jong, E. and Drew, A. and Duran, A. and Ferreira, P. G. and Figueras, P. and Finkel, H. and Fran\c{c}a, T. and Ge, B.-X. and Gu, C. and Helfer, T. and J\"aykk\"a, J. and Joana, C. and Kunesch, M. and Kornet, K. and Lim, E. A. and Muia, F. and Nazari, Z. and Radia, M. and Ripley, J. and Shellard, P. and Sperhake, U. and Traykova, D. and Tunyasuvunakool, S. and Wang, Z. and Widdicombe, J. Y. and Wong, K.},
  title   = {{GRChombo}: An adaptable numerical relativity code for fundamental physics},
  journal = {J. Open Source Softw.},
  volume  = {6},
  pages   = {3703},
  year    = {2021},
  eprint  = {2201.03458},
  archivePrefix = {arXiv}
}

@article{Aurrekoetxea2022CTTK,
  author  = {Aurrekoetxea, J. C. and Clough, K. and Lim, E. A.},
  title   = {{CTTK}: a new method to solve the initial data constraints in numerical relativity},
  journal = {Class. Quant. Grav.},
  volume  = {40},
  pages   = {075003},
  year    = {2023},
  eprint  = {2207.03125},
  archivePrefix = {arXiv}
}

@article{Aurrekoetxea2023,
  author  = {Aurrekoetxea, J. C. and Clough, K. and Muia, F.},
  title   = {Oscillon formation during inflationary preheating with general relativity},
  journal = {Phys. Rev. D},
  volume  = {108},
  pages   = {023501},
  year    = {2023},
  eprint  = {2304.01673},
  archivePrefix = {arXiv}
}

@article{deJong2023,

    author = "de Jong, Eloy and Aurrekoetxea, Josu C. and Lim, Eugene A. and Fran{\c{c}}a, Tiago",
    title = "{Spinning primordial black holes formed during a matter-dominated era}",
    eprint = "2306.11810",
    archivePrefix = "arXiv",
    primaryClass = "astro-ph.CO",
    reportNumber = "KCL-PH-TH/2023-35",
    doi = "10.1088/1475-7516/2023/10/067",
    journal = "JCAP",
    volume = "2023",
    number = "10",
    pages = "067",
    year = "2023"
}

@article{Martin:2019nuw,

    author = "Martin, J{\'e}r{\^o}me and Papanikolaou, Theodoros and Vennin, Vincent",
    title = "{Primordial black holes from the preheating instability in single-field inflation}",
    eprint = "1907.04236",
    archivePrefix = "arXiv",
    primaryClass = "astro-ph.CO",
    doi = "10.1088/1475-7516/2020/01/024",
    journal = "JCAP",
    volume = "2020",
    number = "01",
    pages = "024",
    year = "2020"
}

@article{Zhou:2013tsa,
    author = "Zhou, Shuang-Yong and Copeland, Edmund J. and Easther, Richard and Finkel, Hal and Mou, Zong-Gang and Saffin, Paul M.",
    title = "{Gravitational Waves from Oscillon Preheating}",
    eprint = "1304.6094",
    archivePrefix = "arXiv",
    primaryClass = "astro-ph.CO",
    doi = "10.1007/JHEP10(2013)026",
    journal = "JHEP",
    volume = "10",
    pages = "026",
    year = "2013",
    number = "2013"
}

@article{Lozanov:2019ylm,
    author = "Lozanov, Kaloian D. and Amin, Mustafa A.",
    title = "{Gravitational perturbations from oscillons and transients after inflation}",
    eprint = "1902.06736",
    archivePrefix = "arXiv",
    primaryClass = "astro-ph.CO",
    doi = "10.1103/PhysRevD.99.123504",
    journal = "Phys. Rev. D",
    volume = "99",
    number = "12",
    pages = "123504",
    year = "2019"
}

@article{Antusch:2016con,
    author = "Antusch, Stefan and Cefala, Francesco and Orani, Stefano",
    title = "{Gravitational waves from oscillons after inflation}",
    eprint = "1607.01314",
    archivePrefix = "arXiv",
    primaryClass = "astro-ph.CO",
    doi = "10.1103/PhysRevLett.118.011303",
    journal = "Phys. Rev. Lett.",
    volume = "118",
    number = "1",
    pages = "011303",
    year = "2017",
    note = "[Erratum: Phys.Rev.Lett. 120, 219901 (2018)]"
}

@article{Shafi:2024jig,
    author = "Shafi, Mohammed and Copeland, Edmund J. and Mahbub, Rafid and Mishra, Swagat S. and Basak, Soumen",
    title = "{Formation and decay of oscillons after inflation in the presence of an external coupling. Part I. Lattice simulations}",
    eprint = "2406.00108",
    archivePrefix = "arXiv",
    primaryClass = "hep-ph",
    doi = "10.1088/1475-7516/2024/10/082",
    journal = "JCAP",
    volume = "2024",
    number = "10",
    pages = "082",
    year = "2024"
}

@article{Figueroa:2020rrl,

    author = "Figueroa, Daniel G. and Florio, Adrien and Torrenti, Francisco and Valkenburg, Wessel",
    title = "{The art of simulating the early Universe -- Part I: Integration techniques and canonical cases}",
    eprint = "2006.15122",
    archivePrefix = "arXiv",
    primaryClass = "astro-ph.CO",
    doi = "10.1088/1475-7516/2021/04/035",
    journal = "JCAP",
    volume = "2021",
    number = "04",
    pages = "035",
    year = "2021"
}

@misc{Baeza-Ballesteros:2025tme,
    author = "Baeza-Ballesteros, Jorge and Figueroa, Daniel G. and Florio, Adrien and Lizarraga, Joanes and Loayza, Nicolás and Marschall, Kenneth and Opferkuch, Toby and Stefanek, Ben A. and Torrentí, Francisco and Urio, Ander",
    title = "{The art of simulating the early Universe. Part II. Non-canonical cases and gravitational waves}",
    eprint = "2512.15627",
    archivePrefix = "arXiv",
    primaryClass = "astro-ph.CO",
    reportNumber = "DESY-25-191",
    month = "12",
    year = "2025"
}

@article{Figueroa:2023xmq,
    author = "Figueroa, Daniel G. and Florio, Adrien and Torrenti, Francisco",
    title = "{Present and future of ${\mathcal{C}}$ osmo ${\mathcal{L}}$ attice}",
    eprint = "2312.15056",
    archivePrefix = "arXiv",
    primaryClass = "astro-ph.CO",
    doi = "10.1088/1361-6633/ad616a",
    journal = "Rept. Prog. Phys.",
    volume = "87",
    number = "9",
    pages = "094901",
    year = "2024"
}

@article{Figueroa:2021yhd,
    author = "Figueroa, Daniel G. and Florio, Adrien and Torrenti, Francisco and Valkenburg, Wessel",
    title = "{CosmoLattice: A modern code for lattice simulations of scalar and gauge field dynamics in an expanding universe}",
    eprint = "2102.01031",
    archivePrefix = "arXiv",
    primaryClass = "astro-ph.CO",
    doi = "10.1016/j.cpc.2022.108586",
    journal = "Comput. Phys. Commun.",
    volume = "283",
    pages = "108586",
    year = "2023"
}

@article{Felder:2000hq,
    author = "Felder, Gary N. and Tkachev, Igor",
    title = "{LATTICEEASY: A Program for lattice simulations of scalar fields in an expanding universe}",
    eprint = "hep-ph/0011159",
    archivePrefix = "arXiv",
    reportNumber = "SU-ITP-00-29",
    doi = "10.1016/j.cpc.2008.02.009",
    journal = "Comput. Phys. Commun.",
    volume = "178",
    pages = "929--932",
    year = "2008"
}

@article{Frolov:2008hy,

    author = "Frolov, Andrei V.",
    title = "{DEFROST: A New Code for Simulating Preheating after Inflation}",
    eprint = "0809.4904",
    archivePrefix = "arXiv",
    primaryClass = "hep-ph",
    reportNumber = "SCG-2008-02",
    doi = "10.1088/1475-7516/2008/11/009",
    journal = "JCAP",
    volume = "2008",
    number = "11",
    pages = "009",
    year = "2008",
}

@article{Sainio:2009hm,
    author = "Sainio, Jani",
    title = "{CUDAEASY - a GPU Accelerated Cosmological Lattice Program}",
    eprint = "0911.5692",
    archivePrefix = "arXiv",
    primaryClass = "astro-ph.IM",
    doi = "10.1016/j.cpc.2010.01.002",
    journal = "Comput. Phys. Commun.",
    volume = "181",
    pages = "906--912",
    year = "2010"
}

@article{Bhoonah:2020oov,

    author = "Bhoonah, Amit and Bramante, Joseph and Nerval, Simran and Song, Ningqiang",
    title = "{Gravitational Waves From Dark Sectors, Oscillating Inflatons, and Mass Boosted Dark Matter}",
    eprint = "2008.12306",
    archivePrefix = "arXiv",
    primaryClass = "hep-ph",
    doi = "10.1088/1475-7516/2021/04/043",
    journal = "JCAP",
    volume = "2021",
    number = "04",
    pages = "043",
    year = "2021",
}

@article{Huang:2011gf,
    author = "Huang, Zhiqi",
    title = "{The Art of Lattice and Gravity Waves from Preheating}",
    eprint = "1102.0227",
    archivePrefix = "arXiv",
    primaryClass = "astro-ph.CO",
    doi = "10.1103/PhysRevD.83.123509",
    journal = "Phys. Rev. D",
    volume = "83",
    pages = "123509",
    year = "2011"
}

@article{del-Corral:2025fca,
    author = "del-Corral, Daniel and Kumar, K. Sravan and Marto, Jo{\~a}o",
    title = "{Gravitational waves from primordial black hole dominance: The effect of inflaton decay rate}",
    eprint = "2504.05875",
    archivePrefix = "arXiv",
    primaryClass = "astro-ph.CO",
    doi = "10.1016/j.dark.2025.101991",
    journal = "Phys. Dark Univ.",
    volume = "49",
    pages = "101991",
    year = "2025"
}

@article{Martin:2020fgl,

    author = "Martin, J{\'e}r{\^o}me and Papanikolaou, Theodoros and Pinol, Lucas and Vennin, Vincent",
    title = "{Metric preheating and radiative decay in single-field inflation}",
    eprint = "2002.01820",
    archivePrefix = "arXiv",
    primaryClass = "astro-ph.CO",
    doi = "10.1088/1475-7516/2020/05/003",
    journal = "JCAP",
    volume = "2020",
    number = "05",
    pages = "003",
    year = "2020"
}

@article{del-Corral:2023apl,

    author = "del-Corral, Daniel and Gondolo, Paolo and Kumar, K. Sravan and Marto, Jo{\~a}o",
    title = "{Revisiting primordial black holes formation from preheating instabilities: the case of Starobinsky inflation}",
    eprint = "2311.02754",
    archivePrefix = "arXiv",
    primaryClass = "astro-ph.CO",
    doi = "10.1088/1475-7516/2025/02/009",
    journal = "JCAP",
    volume = "2025",
    number = "02",
    pages = "009",
    year = "2025"
}

@article{del-Corral:2024vcm,
    author = "del-Corral, Daniel",
    title = "{Self-resonance during preheating: The case of {\ensuremath{\alpha}}-attractor models}",
    eprint = "2406.04017",
    archivePrefix = "arXiv",
    primaryClass = "hep-th",
    doi = "10.1016/j.aop.2024.169824",
    journal = "Annals Phys.",
    volume = "470",
    pages = "169824",
    year = "2024"
}

@article{Hertzberg:2014iza,
    author = "Hertzberg, Mark P. and Karouby, Johanna and Spitzer, William G. and Becerra, Juana C. and Li, Lanqing",
    title = "{Theory of self-resonance after inflation. I. Adiabatic and isocurvature Goldstone modes}",
    eprint = "1408.1396",
    archivePrefix = "arXiv",
    primaryClass = "hep-th",
    reportNumber = "MIT-CTP-4571",
    doi = "10.1103/PhysRevD.90.123528",
    journal = "Phys. Rev. D",
    volume = "90",
    pages = "123528",
    year = "2014"
}

@article{Hertzberg:2014jza,
    author = "Hertzberg, Mark P. and Karouby, Johanna and Spitzer, William G. and Becerra, Juana C. and Li, Lanqing",
    title = "{Theory of self-resonance after inflation. II. Quantum mechanics and particle-antiparticle asymmetry}",
    eprint = "1408.1398",
    archivePrefix = "arXiv",
    primaryClass = "hep-th",
    reportNumber = "MIT-CTP-4573",
    doi = "10.1103/PhysRevD.90.123529",
    journal = "Phys. Rev. D",
    volume = "90",
    pages = "123529",
    year = "2014"
}

@article{Jedamzik:2010dq,

    author = "Jedamzik, Karsten and Lemoine, Martin and Martin, Jerome",
    title = "{Collapse of Small-Scale Density Perturbations during Preheating in Single Field Inflation}",
    eprint = "1002.3039",
    archivePrefix = "arXiv",
    primaryClass = "astro-ph.CO",
    doi = "10.1088/1475-7516/2010/09/034",
    journal = "JCAP",
    volume = "2010",
    number = "09",
    pages = "034",
    year = "2010"
}

@article{Sfakianakis:2018lzf,
    author = "Sfakianakis, Evangelos I. and van de Vis, Jorinde",
    title = "{Preheating after Higgs Inflation: Self-Resonance and Gauge boson production}",
    eprint = "1810.01304",
    archivePrefix = "arXiv",
    primaryClass = "hep-ph",
    reportNumber = "Nikhef-2018-044",
    doi = "10.1103/PhysRevD.99.083519",
    journal = "Phys. Rev. D",
    volume = "99",
    number = "8",
    pages = "083519",
    year = "2019"
}

@article{Ballesteros:2024eee,
    author = "Ballesteros, Guillermo and Iguaz Juan, Joaquim and Serpico, Pasquale D. and Taoso, Marco",
    title = "{Primordial black hole formation from self-resonant preheating?}",
    eprint = "2406.09122",
    archivePrefix = "arXiv",
    primaryClass = "astro-ph.CO",
    doi = "10.1103/PhysRevD.111.083521",
    journal = "Phys. Rev. D",
    volume = "111",
    number = "8",
    pages = "083521",
    year = "2025"
}

@article{Lozanov:2019jff,

    author = "Lozanov, Kaloian D. and Amin, Mustafa A.",
    title = "{GFiRe{\textemdash}Gauge Field integrator for Reheating}",
    eprint = "1911.06827",
    archivePrefix = "arXiv",
    primaryClass = "astro-ph.CO",
    doi = "10.1088/1475-7516/2020/04/058",
    journal = "JCAP",
    volume = "2020",
    number = "04",
    pages = "058",
    year = "2020"
}

@article{Felder:2007kat,
    author = "Felder, Gary N.",
    title = "{CLUSTEREASY: A program for lattice simulations of scalar fields in an expanding universe on parallel computing clusters}",
    eprint = "0712.0813",
    archivePrefix = "arXiv",
    primaryClass = "hep-ph",
    doi = "10.1016/j.cpc.2008.06.002",
    journal = "Comput. Phys. Commun.",
    volume = "179",
    pages = "604--606",
    year = "2008"
}

@article{Sainio:2012mw,

    author = "Sainio, J.",
    title = "{PyCOOL - a Cosmological Object-Oriented Lattice code written in Python}",
    eprint = "1201.5029",
    archivePrefix = "arXiv",
    primaryClass = "astro-ph.IM",
    doi = "10.1088/1475-7516/2012/04/038",
    journal = "JCAP",
    volume = "2012",
    number = "04",
    pages = "038",
    year = "2012",
}

@article{Child:2013ria,
    author = "Child, Hillary L. and Giblin, Jr, John T. and Ribeiro, Raquel H. and Seery, David",
    title = "{Preheating with Non-Minimal Kinetic Terms}",
    eprint = "1305.0561",
    archivePrefix = "arXiv",
    primaryClass = "astro-ph.CO",
    doi = "10.1103/PhysRevLett.111.051301",
    journal = "Phys. Rev. Lett.",
    volume = "111",
    pages = "051301",
    year = "2013"
}

@article{Easther:2010qz,

    author = "Easther, Richard and Finkel, Hal and Roth, Nathaniel",
    title = "{PSpectRe: A Pseudo-Spectral Code for (P)reheating}",
    eprint = "1005.1921",
    archivePrefix = "arXiv",
    primaryClass = "astro-ph.CO",
    doi = "10.1088/1475-7516/2010/10/025",
    journal = "JCAP",
    volume = "2010",
    number = "10",
    pages = "025",
    year = "2010",
}

@article{Giblin:2019nuv,
    author = "Giblin, John T. and Tishue, Avery J.",
    title = "{Preheating in Full General Relativity}",
    eprint = "1907.10601",
    archivePrefix = "arXiv",
    primaryClass = "gr-qc",
    doi = "10.1103/PhysRevD.100.063543",
    journal = "Phys. Rev. D",
    volume = "100",
    number = "6",
    pages = "063543",
    year = "2019"
}

@article{Turner:1983he,
    author = "Turner, Michael S.",
    title = "{Coherent Scalar Field Oscillations in an Expanding Universe}",
    reportNumber = "EFI-83-29-CHICAGO",
    doi = "10.1103/PhysRevD.28.1243",
    journal = "Phys. Rev. D",
    volume = "28",
    pages = "1243",
    year = "1983"
}

@article{Press:1989id,
    author = "Press, William H. and Ryden, Barbara S. and Spergel, David N.",
    title = "{Single Mechanism for Generating Large Scale Structure and Providing Dark Missing Matter}",
    reportNumber = "CFA-3031",
    doi = "10.1103/PhysRevLett.64.1084",
    journal = "Phys. Rev. Lett.",
    volume = "64",
    pages = "1084",
    year = "1990"
}

@article{Kolb:1993hw,
    author = "Kolb, Edward W. and Tkachev, Igor I.",
    title = "{Nonlinear axion dynamics and formation of cosmological pseudosolitons}",
    eprint = "astro-ph/9311037",
    archivePrefix = "arXiv",
    reportNumber = "FERMILAB-PUB-93-335-A",
    doi = "10.1103/PhysRevD.49.5040",
    journal = "Phys. Rev. D",
    volume = "49",
    pages = "5040--5051",
    year = "1994"
}

@article{Kolb:1994fi,
    author = "Kolb, Edward W. and Tkachev, Igor I.",
    title = "{Large amplitude isothermal fluctuations and high density dark matter clumps}",
    eprint = "astro-ph/9403011",
    archivePrefix = "arXiv",
    reportNumber = "FERMILAB-PUB-94-055-A",
    doi = "10.1103/PhysRevD.50.769",
    journal = "Phys. Rev. D",
    volume = "50",
    pages = "769--773",
    year = "1994"
}

@article{Easther:2006vd,
    author = "Easther, Richard and Giblin, Jr., John T. and Lim, Eugene A.",
    title = "{Gravitational Wave Production At The End Of Inflation}",
    eprint = "astro-ph/0612294",
    archivePrefix = "arXiv",
    doi = "10.1103/PhysRevLett.99.221301",
    journal = "Phys. Rev. Lett.",
    volume = "99",
    pages = "221301",
    year = "2007"
}

@article{Easther:2007vj,
    author = "Easther, Richard and Giblin, John T. and Lim, Eugene A.",
    title = "{Gravitational Waves From the End of Inflation: Computational Strategies}",
    eprint = "0712.2991",
    archivePrefix = "arXiv",
    primaryClass = "astro-ph",
    doi = "10.1103/PhysRevD.77.103519",
    journal = "Phys. Rev. D",
    volume = "77",
    pages = "103519",
    year = "2008"
}

@article{Grutkoski:2025ygs,
    author = "Grutkoski, Ryn and Macpherson, Hayley J. and Giblin, Jr., John T. and Frieman, Joshua",
    title = "{Effect of nonlinear gravity on the cosmological background during preheating}",
    eprint = "2504.08939",
    archivePrefix = "arXiv",
    primaryClass = "astro-ph.CO",
    doi = "10.1103/3sgw-48pk",
    journal = "Phys. Rev. D",
    volume = "112",
    number = "2",
    pages = "023541",
    year = "2025"
}

@article{Adshead:2023mvt,

    author = "Adshead, Peter and Giblin, John T. and Grutkoski, Ryn and Weiner, Zachary J.",
    title = "{Gauge preheating with full general relativity}",
    eprint = "2311.01504",
    archivePrefix = "arXiv",
    primaryClass = "astro-ph.CO",
    doi = "10.1088/1475-7516/2024/03/017",
    journal = "JCAP",
    volume = "2024",
    number = "03",
    pages = "017",
    year = "2024"
}

@misc{Baumgarte:2026igz,
    author = "Baumgarte, Thomas W. and Clough, Katy and Gerhardinger, Mary and Giblin, John T. and Miller, Amanda",
    title = "{Primordial Black Holes in a Radiation-Dominated Universe}",
    eprint = "2606.30641",
    archivePrefix = "arXiv",
    primaryClass = "astro-ph.CO",
    month = "6",
    year = "2026"
}

@article{Starobinsky:1980te,
    author = "Starobinsky, Alexei A.",
    editor = "Khalatnikov, I. M. and Mineev, V. P.",
    title = "{A New Type of Isotropic Cosmological Models Without Singularity}",
    doi = "10.1016/0370-2693(80)90670-X",
    journal = "Phys. Lett. B",
    volume = "91",
    pages = "99--102",
    year = "1980"
}

@article{Kallosh:2013daa,

    author = "Kallosh, Renata and Linde, Andrei",
    title = "{Multi-field Conformal Cosmological Attractors}",
    eprint = "1309.2015",
    archivePrefix = "arXiv",
    primaryClass = "hep-th",
    doi = "10.1088/1475-7516/2013/12/006",
    journal = "JCAP",
    volume = "2013",
    number = "12",
    pages = "006",
    year = "2013"
}

@article{Kallosh:2013hoa,

    author = "Kallosh, Renata and Linde, Andrei",
    title = "{Universality Class in Conformal Inflation}",
    eprint = "1306.5220",
    archivePrefix = "arXiv",
    primaryClass = "hep-th",
    doi = "10.1088/1475-7516/2013/07/002",
    journal = "JCAP",
    volume = "2013",
    number = "07",
    pages = "002",
    year = "2013"
}

@article{Kaiser:2013sna,
    author = "Kaiser, David I. and Sfakianakis, Evangelos I.",
    title = "{Multifield Inflation after Planck: The Case for Nonminimal Couplings}",
    eprint = "1304.0363",
    archivePrefix = "arXiv",
    primaryClass = "astro-ph.CO",
    reportNumber = "PREPRINT-MIT-CTP-4451",
    doi = "10.1103/PhysRevLett.112.011302",
    journal = "Phys. Rev. Lett.",
    volume = "112",
    number = "1",
    pages = "011302",
    year = "2014"
}

@article{Kallosh:2015lwa,
    author = "Kallosh, Renata and Linde, Andrei",
    title = "{Planck, LHC, and $\alpha$-attractors}",
    eprint = "1502.07733",
    archivePrefix = "arXiv",
    primaryClass = "astro-ph.CO",
    doi = "10.1103/PhysRevD.91.083528",
    journal = "Phys. Rev. D",
    volume = "91",
    pages = "083528",
    year = "2015"
}

@article{Shibata:1995,
  title = {Evolution of three-dimensional gravitational waves: Harmonic slicing case},
  author = {Shibata, Masaru and Nakamura, Takashi},
  journal = {Phys. Rev. D},
  volume = {52},
  issue = {10},
  pages = {5428--5444},
  numpages = {0},
  year = {1995},
  month = {Nov},
  publisher = {American Physical Society},
  doi = {10.1103/PhysRevD.52.5428},
  url = {https://link.aps.org/doi/10.1103/PhysRevD.52.5428}
}

@article{Baumgarte:1998te,
    author = "Baumgarte, Thomas W. and Shapiro, Stuart L.",
    title = "{On the numerical integration of Einstein's field equations}",
    eprint = "gr-qc/9810065",
    archivePrefix = "arXiv",
    doi = "10.1103/PhysRevD.59.024007",
    journal = "Phys. Rev. D",
    volume = "59",
    pages = "024007",
    year = "1998"
}

@article{Nakamura:1987zz,
    author = "Nakamura, T. and Oohara, K. and Kojima, Y.",
    title = "{General Relativistic Collapse to Black Holes and Gravitational Waves from Black Holes}",
    doi = "10.1143/PTPS.90.1",
    journal = "Prog. Theor. Phys. Suppl.",
    volume = "90",
    pages = "1--218",
    year = "1987"
}

@article{Arnowitt:1962hi,
    author = "Arnowitt, Richard L. and Deser, Stanley and Misner, Charles W.",
    title = "{The Dynamics of general relativity}",
    eprint = "gr-qc/0405109",
    archivePrefix = "arXiv",
    doi = "10.1007/s10714-008-0661-1",
    journal = "Gen. Rel. Grav.",
    volume = "40",
    pages = "1997--2027",
    year = "2008"
}

@article{del-Corral:2025lcp,
    author = "del-Corral, Daniel and Gondolo, Paolo and Kumar, K. Sravan and Marto, Jo{\~a}o",
    title = "{Primordial black holes through preheating instabilities in {\ensuremath{\alpha}}-attractor models}",
    eprint = "2505.17790",
    archivePrefix = "arXiv",
    primaryClass = "astro-ph.CO",
    doi = "10.1103/v9jt-1zpm",
    journal = "Phys. Rev. D",
    volume = "114",
    pages = "023561",
    year = "2026"
}

@article{Alic:2011gg,
    author = "Alic, Daniela and Bona-Casas, Carles and Bona, Carles and Rezzolla, Luciano and Palenzuela, Carlos",
    title = "{Conformal and covariant formulation of the Z4 system with constraint-violation damping}",
    eprint = "1106.2254",
    archivePrefix = "arXiv",
    primaryClass = "gr-qc",
    doi = "10.1103/PhysRevD.85.064040",
    journal = "Phys. Rev. D",
    volume = "85",
    pages = "064040",
    year = "2012"
}

@article{Pretorius:2004jg,
    author = "Pretorius, Frans",
    title = "{Numerical relativity using a generalized harmonic decomposition}",
    eprint = "gr-qc/0407110",
    archivePrefix = "arXiv",
    doi = "10.1088/0264-9381/22/2/014",
    journal = "Class. Quant. Grav.",
    volume = "22",
    pages = "425--452",
    year = "2005"
}

@article{Bona:2003fj,
    author = "Bona, C. and Ledvinka, T. and Palenzuela, C. and Zacek, M.",
    title = "{General covariant evolution formalism for numerical relativity}",
    eprint = "gr-qc/0302083",
    archivePrefix = "arXiv",
    doi = "10.1103/PhysRevD.67.104005",
    journal = "Phys. Rev. D",
    volume = "67",
    pages = "104005",
    year = "2003"
}

@article{Weyhausen:2011cg,
    author = "Weyhausen, Andreas and Bernuzzi, Sebastiano and Hilditch, David",
    title = "{Constraint damping for the Z4c formulation of general relativity}",
    eprint = "1107.5539",
    archivePrefix = "arXiv",
    primaryClass = "gr-qc",
    doi = "10.1103/PhysRevD.85.024038",
    journal = "Phys. Rev. D",
    volume = "85",
    pages = "024038",
    year = "2012"
}

@article{Mahbub:2023faw,
    author = "Mahbub, Rafid and Mishra, Swagat S.",
    title = "{Oscillon formation from preheating in asymmetric inflationary potentials}",
    eprint = "2303.07503",
    archivePrefix = "arXiv",
    primaryClass = "astro-ph.CO",
    doi = "10.1103/PhysRevD.108.063524",
    journal = "Phys. Rev. D",
    volume = "108",
    number = "6",
    pages = "063524",
    year = "2023"
}

@article{Amin:2011hj,
    author = "Amin, Mustafa A. and Easther, Richard and Finkel, Hal and Flauger, Raphael and Hertzberg, Mark P.",
    title = "{Oscillons After Inflation}",
    eprint = "1106.3335",
    archivePrefix = "arXiv",
    primaryClass = "astro-ph.CO",
    doi = "10.1103/PhysRevLett.108.241302",
    journal = "Phys. Rev. Lett.",
    volume = "108",
    pages = "241302",
    year = "2012"
}

@article{del-Corral:2025fzz,
    author = "del-Corral, Daniel and Gondolo, Paolo and Kumar, K. Sravan and Marto, Jo{\~a}o",
    title = "{Scalar-induced gravitational waves from self-resonant preheating in {\ensuremath{\alpha}}-attractor models}",
    eprint = "2504.17602",
    archivePrefix = "arXiv",
    primaryClass = "astro-ph.CO",
    doi = "10.1103/q69z-gw6l",
    journal = "Phys. Rev. D",
    volume = "113",
    number = "2",
    pages = "023526",
    year = "2026"
}

@book{Baumgarte:2021skc,
    author = "Baumgarte, Thomas W. and Shapiro, Stuart L.",
    title = "{Numerical Relativity: Starting from Scratch}",
    doi = "10.1017/9781108933445",
    isbn = "978-1-108-93344-5, 978-1-108-84411-6, 978-1-108-92825-0",
    publisher = "Cambridge University Press",
    month = "2",
    year = "2021"
}

@book{Baumgarte:2010ndz,
    author = "Baumgarte, Thomas W. and Shapiro, Stuart L.",
    title = "{Numerical Relativity: Solving Einstein's Equations on the Computer}",
    doi = "10.1017/CBO9781139193344",
    publisher = "Cambridge University Press",
    year = "2010"
}

@article{Gourgoulhon:2007ue,
    author        = {Gourgoulhon, Eric},
    title         = {3+1 Formalism and Bases of Numerical Relativity},
    eprint        = {gr-qc/0703035},
    archivePrefix = {arXiv},
    primaryClass  = {gr-qc},
    year          = {2007},
    journal = {preprint}
}

@article{Whitt:1984pd,
  author  = {Whitt, Brian},
  title   = {Fourth-Order Gravity as General Relativity Plus Matter},
  journal = {Phys. Lett. B},
  volume  = {145},
  pages   = {176--178},
  year    = {1984},
  doi     = {10.1016/0370-2693(84)90332-0}
}

@article{Maeda:1988ab,
  author  = {Maeda, Kei-ichi},
  title   = {Towards the Einstein-Hilbert Action via Conformal Transformation},
  journal = {Phys. Rev. D},
  volume  = {39},
  pages   = {3159},
  year    = {1989},
  doi     = {10.1103/PhysRevD.39.3159}
}

\end{document}